*white paper on*

# Pluto System Follow On Missions: Background, Rationale, and New Mission Recommendations

*primary authors*

**Stuart J. Robbins** Southwest Research Institute — stuart@boulder.swri.edu*

**S. Alan Stern** Southwest Research Institute — alan@boulder.swri.edu

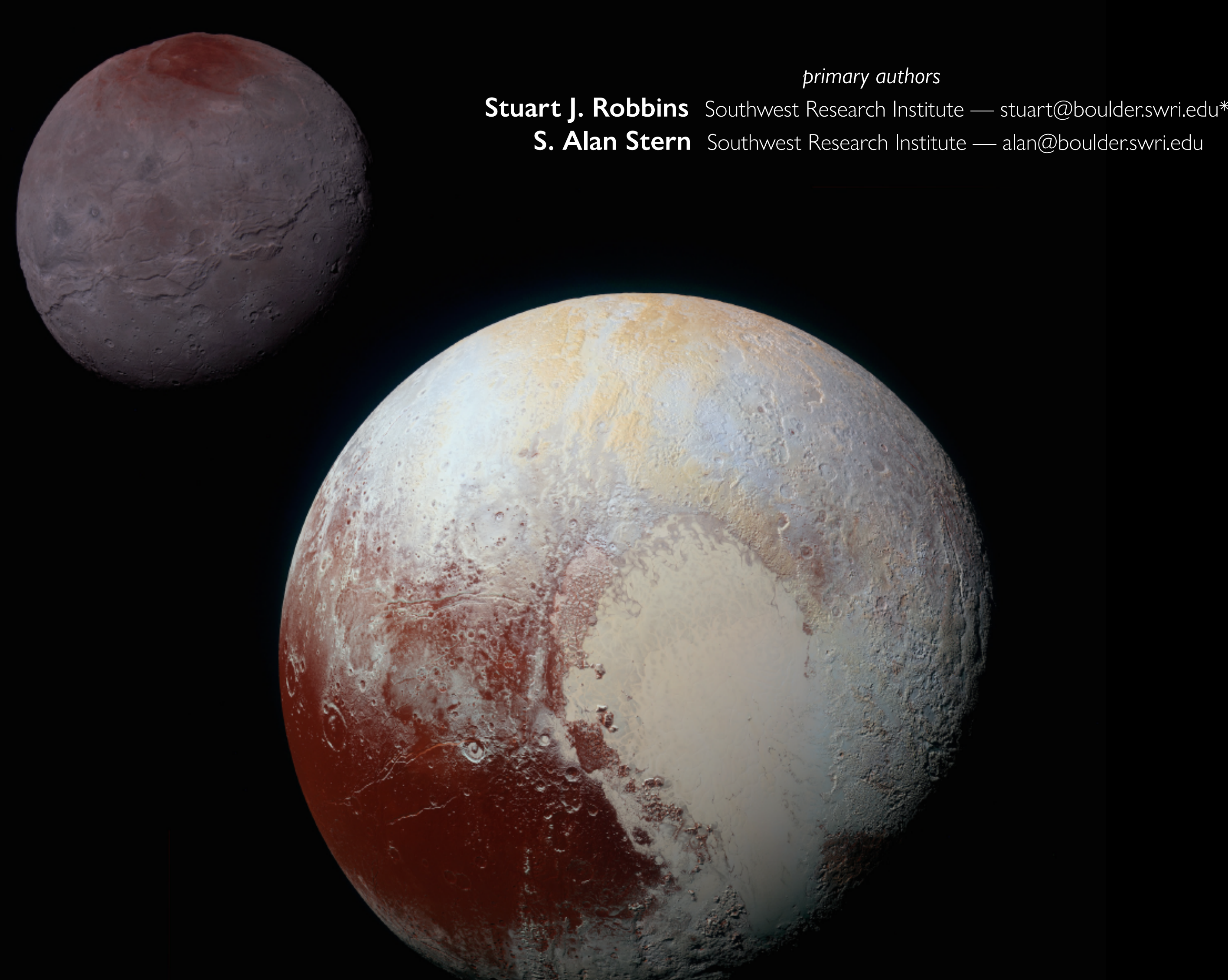

*co-authors (alphabetical)*

**Richard Binzel** (MIT), **Will Grundy** (Lowell Observatory), **Doug Hamilton** (University of Maryland), **Rosaly Lopes** (NASA-JPL), **Bill McKinnon** (Washington University in St. Louis), and **Cathy Olkin** (Southwest Research Institute)

*endorsers / co-signers*

Caitlin Ahrens, Michelle T. Bannister, Daniel C. Berman, Tanguy Bertrand, Ross Beyer, Maitrayee Bose, Paul Byrne, Robert Chancia, Dale Cruikshank, Adeene Denton, Rajani Dhingra, Cynthia Dinwiddie, David Dunham, Alissa Earle, Edith C. Fayolle, Christopher Glein, Cesare Grava, Aurelie Guilbert, Mark Gurwell, Jennifer Hanley, Jason Hofgartner, Bryan Holler, Mihaly Horanyi, Sona Hosseini, Carly Howett, Reggie Hudson, James Keane, Akos Kereszturi, Eduard Kuznetsov, Alice Lucchetti, Renu Malhotra, Kathleen Mandt, Tim I. Michaels, Laurent Montesi, Jeff Moore, Jessica Noviello, Benoît Noyelles, Maurizio Pajola, Silvia Protopapa, Lynnae Quick, Jani Radebaugh, Gustavo Benedetti Rossi, Emilie Royer, Kirby Runyon, Pablo Santos Sanz, Paul Schenk, Jennifer Scully, Amanda Sickafoose, Kelsi Singer, Timothy Stubbs, Mark Sykes, Laurence Trafton, Orkan Umurhan, Anne Verbiscer, Larry Wasserman, Oliver L. White, Teresa Wong, Leslie Young, and Amanda Zangari

*Montage of Pluto and Charon, as encountered near the closest approach of New Horizons. Color is enhanced, but the approximate sizes of the two bodies and relative brightnesses are accurate. Credit: NASA / JHU-APL / SwRI / Alex H. Parker*

## Key Findings

- The Pluto system is extremely diverse, despite residing >40 A.U. from the Sun, and shows a variety of features and dynamics that warrant further study.
- The ability for resolved time-variability studies, global mapping, second-generation mission instrumentation such as ground-penetrating radar, and *in situ* measurements of Pluto's exosphere, all argue for an enhanced follow on mission to the Pluto system.

## 1. Introduction

The first exploration of Pluto was motivated by (i) the many intriguing aspects of this body, its atmosphere, and its giant impact binary-planet formation; as well as (ii) the scientific desire to initiate the reconnaissance of the newly-discovered population of dwarf planets in the Kuiper Belt (e.g., Belton *et al.* 2003). That exploration took place in the form of a single spacecraft flyby that yielded an impressive array of exciting results that have transformed our understanding of this world and its satellites (Stern *et al.,* 2015), and which opened our eyes to the exciting nature of the dwarf planet population of the Kuiper Belt. From Pluto's five-object satellite system, to its hydrocarbon haze-laden $N_2$-$CH_4$-CO atmosphere, to its variegated distribution of surface volatiles, to its wide array of geologic expressions that include extensive glaciation and suspected cryovolcanoes, plus the tantalizing possibility of an interior ocean, the Pluto system has proven to be as complex as larger terrestrial bodies like Mars (e.g., Moore *et al.*, 2016; Gladstone *et al.*, 2016; Grundy *et al.*, 2016; Olkin *et al.*, 2017; Stern *et al.,* 2018), and it begs for future exploration.

Owing to Pluto's high obliquity (and consequently, current-epoch southern hemisphere polar winter darkness) and the single spacecraft nature of the *New Horizons* flyby, only ~40% of Pluto and its binary satellite, Charon, could be mapped at high pixel scales (better than 10 km/pix). Additionally, due to their distances from *New Horizons* at closest approach, none of Pluto's small moons could be studied at high resolution during the flyby. Furthermore, studies of the time variability of atmospheric, geologic, and surface-atmosphere interactions cannot be practically made by additional flybys, and they cannot be made from Earth-based observations.

We find that these limitations, combined with Pluto's many important, open scientific questions, strongly motivate a Pluto System follow on orbiter mission.

## 2. The Case for a Pluto System Follow On Mission

The reasons to return to Pluto are multifold, as we summarize here. We begin with Pluto's surface and interior, then go on to its atmosphere, its satellites, and finally to Pluto's context in the Kuiper Belt.

### 2.1 Geological and Compositional Diversity and Geophysical Processes on Pluto

The *New Horizons* encounter revealed evidence for a world of ongoing, diverse geological activity, similar in extent and variety to Mars (e.g., Moore *et al.*, 2016; Stern *et al.*, 2018). While certain aspects of Pluto's complex geology were predicted (e.g., Moore *et al.*, 2015), the diversity of activity and novel processes were not anticipated. These include: (i) ancient and ongoing $N_2$-ice glacial activity, (ii) convective overturn in a vast, kilometers-thick, $N_2$-rich ice sheet contained in an ancient basin, (iii) multiple large, potentially cryovolcanic constructs, (iv) aligned blades of methane ice hundreds of meters tall and stretching across hundreds of kilometers, (v) an extreme range of surface ages based on crater spatial densities, and (vi) evidence for a surviving cold ocean under Pluto's surface (e.g., Howard *et al.*, 2017; McKinnon *et al.*, 2016; Moore *et al.*, 2018; Nimmo *et al.*, 2016). Evidence for these and other features such as regional compositional and color diversity (e.g., Grundy *et al.*, 2016; Protopapa *et al.*, 2017) resulted from the high-resolution observations made by *New Horizons* during its brief flyby in 2015.

However, *New Horizons* only studied about 40% of Pluto's surface in detail; the rest was observed either at very low resolution on the anti-encounter hemisphere or obscured by darkness in the southern regions due to winter. Further, *New Horizons* only studied Pluto at high resolution for a period of <24 hours, and it carried a powerful but limited suite of first reconnaissance instrument capabilities. To understand the surface of Pluto, there is a clear need to: (i) image all remaining terrains (*e.g.*, using Charon light or active sensors in polar darkened terrains); (ii) obtain higher resolution geological and compositional maps; (iii) obtain datasets from new kinds of instruments such as ground penetrating radars, mass spectrometers, thermal mappers, and altimeters; (iv) obtain well-resolved gravity measurements; and (v) study time-dependent phenomena.

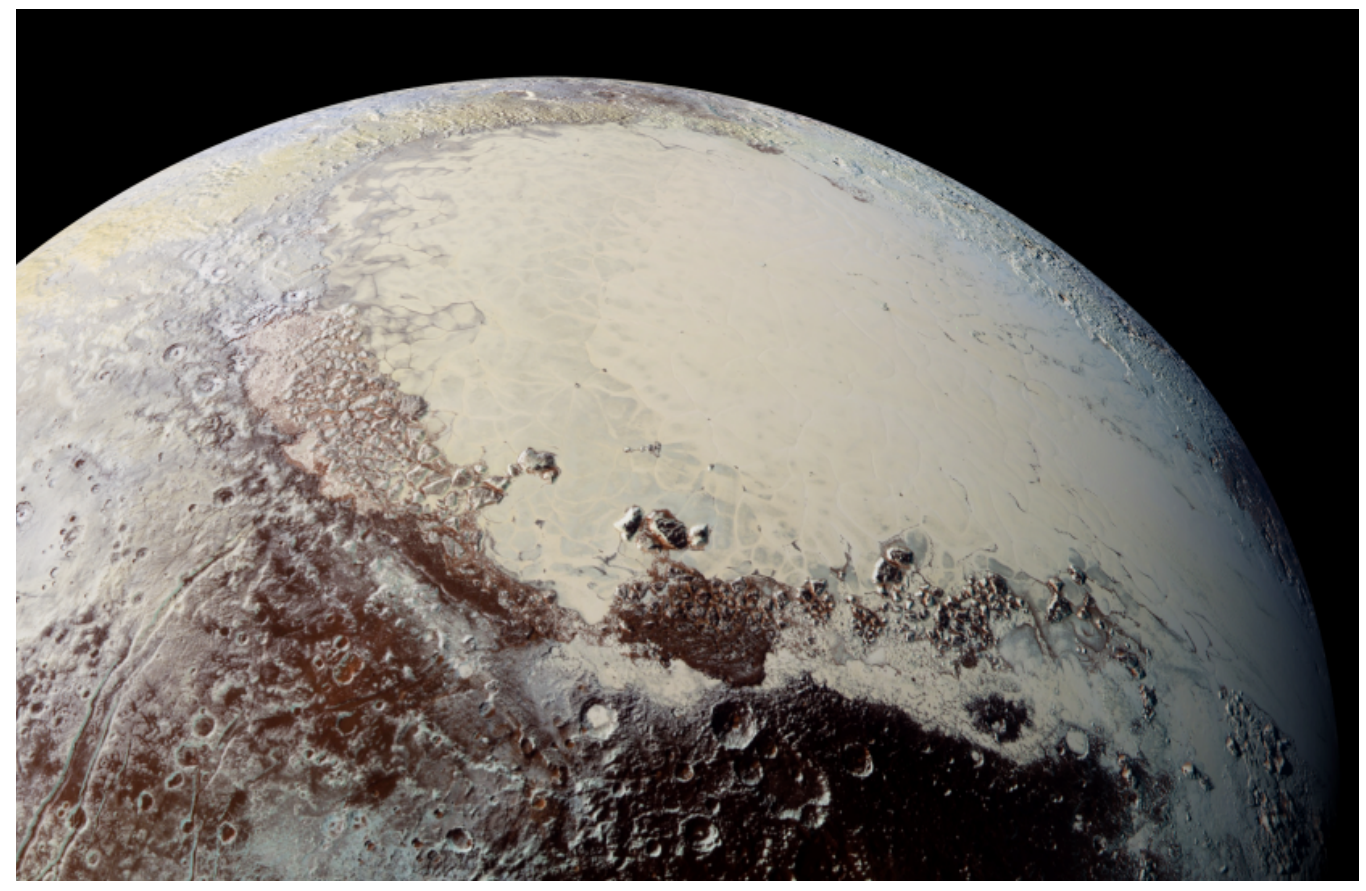

*3D rendering of Pluto in enhanced color showing the vast Sputnik Planitia $N_2$-dominated glacier and surrounding mountain ranges. Credit: NASA / JHU-APL / SwRI / Stuart J. Robbins*

Major questions that must be addressed now include details of Pluto's interior structure which could be constrained by global gravity measurements, such as, does there exist a liquid subsurface ocean today? And, if there is an ocean, how deep is it and what is its extent and composition? When were Pluto's cryovolcanoes active and to what extent? How were the bladed terrains constructed? What caused the formation of Pluto's giant rift system and other tectonic features? What powers Pluto's ongoing geological activity?

## 2.2 Atmosphere, Climate, and Interactions with Pluto's Surface

Pluto's atmosphere and surface function as an interconnected system (*e.g.*, Gladstone *et al.*, 2016; Stern *et al.*, 2018). Surface composition and topography, together, interact with the atmosphere because the atmosphere is supported by vapor pressure equilibrium. Therefore, both the distribution of volatile ices on Pluto's surface and its atmospheric pressure are dynamic and respond to the received insolation. Clearly, one cannot understand either the distribution of surface volatiles or the atmospheric structure in isolation: they are dependent on each other, and they are also dependent on both short-term (diurnal) and long-term (orbital) factors and timescales.

Observations from ground-based stellar occultations have shown that the atmospheric pressure on Pluto increased by a factor of three between 1988 and 2015 as Pluto receded from the Sun (Sicardy *et al.*, 2016). Now that topography and the distribution of volatiles on Pluto are known—at least for the encounter hemisphere (Grundy *et al.*, 2016)—detailed global circulation models can be run. Bertrand & Forget (2016) have been able to numerically replicate the observations from *New Horizons* and the increase in atmospheric pressure detected from ground-based stellar occultations. These models predict significant changes in the distribution of volatiles even on timescales of a terrestrial decade (just 4% of Pluto's orbital period), including the disappearance of mid- and high-latitude frost bands.

Another striking atmospheric discovery made by *New Horizons* was the extent of haze in Pluto's atmosphere. Yet, owing to it being a flyby, *New Horizons* could not study the dynamics or formation of this haze in any significant detail, nor could it see responses in haze production and destruction to diurnal, orbital, seasonal, and solar forcing (as, *e.g.*, *Cassini* was able to do at Titan).

*New Horizons* also lacked the capability to make in-situ atmospheric measurements of composition, haze-size particle frequency distributions, or to study atmospheric dynamics. To accomplish these objectives, and to observe volatile transport and the detailed evolution of the atmosphere (and its escape rate) due to solar cycle and orbital/seasonal effects, requires a new mission with new measurement capabilities, and the ability to remain at Pluto for several terrestrial years. New atmospheric capabilities that are warranted include in-situ mass spectroscopy, nephelometry, and ion/electron density measurements.

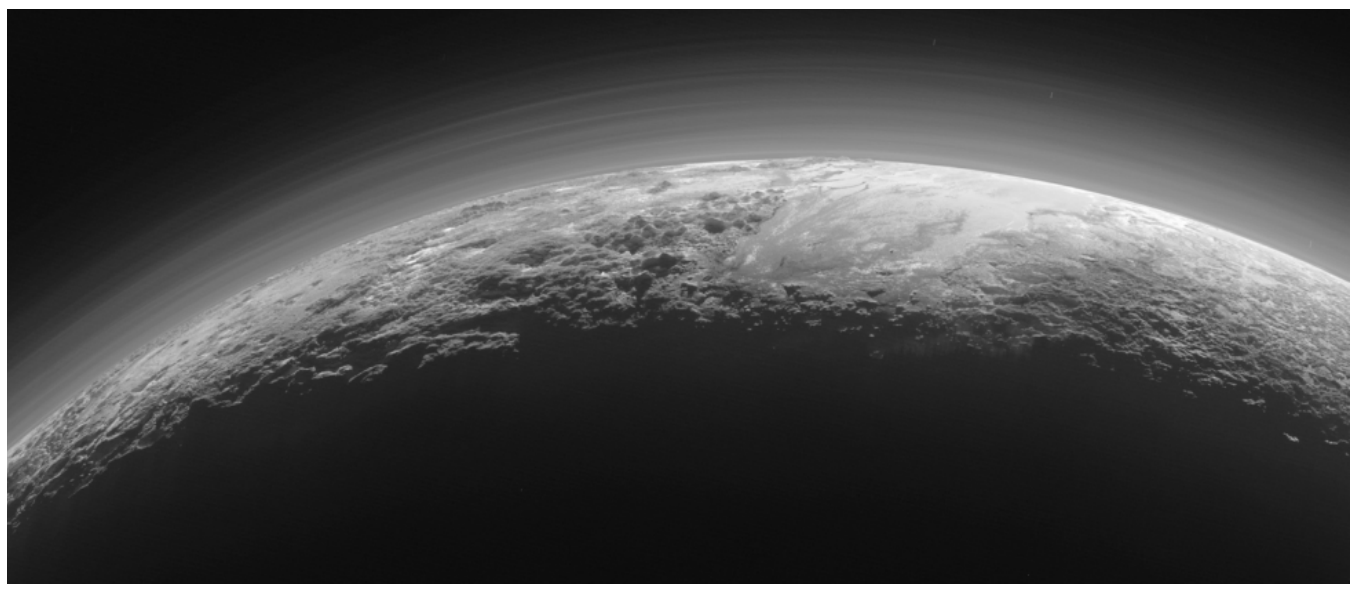

*Pluto's steep topography with characteristic 3–5 km amplitudes and laterally organized atmospheric hazes extending >200 km in altitude. Credit: NASA / JHU-APL / SwRI*

## 2.3 Charon, Pluto's Large Satellite

Charon is comparable in size to the mid-sized Saturnian and Uranian satellites, and it shares with them a cratered, water-ice rich surface. But, it also stands out in ways that may provide insights into stages of evolution common to icy worlds. Ancient terrains are better preserved at Charon owing to reduced impact, radiation, and thermal damage/processing in the Kuiper Belt relative to the regular satellites near the giant planets. Therefore, Charon can provide unique insight into the evolution of icy ocean worlds, particularly when compared to icy satellites of the ice and gas giants. Charon can also provide key insights into binary planets and also, importantly, Earth-Moon system, formation—an objective that cannot be accomplished with any closer system.

*New Horizons* images showed that Charon's surface geology and interior geophysics present important challenges that require future exploration to understand (e.g., Olkin *et al.*, 2017; Stern *et al.*, 2018). For example, one striking feature of Charon's encounter (sub-Pluto) hemisphere is the dichotomy between Vulcan Planitia, the smoother equatorial plains, and the widespread rougher terrains to its north (Moore *et al.*, 2016; Robbins *et al.*, 2017, 2019). Perhaps Vulcan Planitia was formed by the eruption of vast amounts of internal water "magma," forced to the surface upon expansion of a freezing internal liquid water ocean (e.g., Beyer *et al.*, 2017, 2018; Robbins *et al.*, 2019). This last melt could have been especially rich in antifreeze substances such as $NH_3$ that would increase its viscosity (Kargel *et al.*,1991). A related scenario involves foundering of blocks of an ancient icy crust, as the distinctive morphology of Kubrick, Clarke, and Butler Montes (the mountains in depressions) and that of a similar-scale cavity with no mountain are suggestive of blocks having been submerged into an extremely viscous fluid or slurry (e.g., Beyer *et al.*, 2018; Robbins *et al.*, 2019).

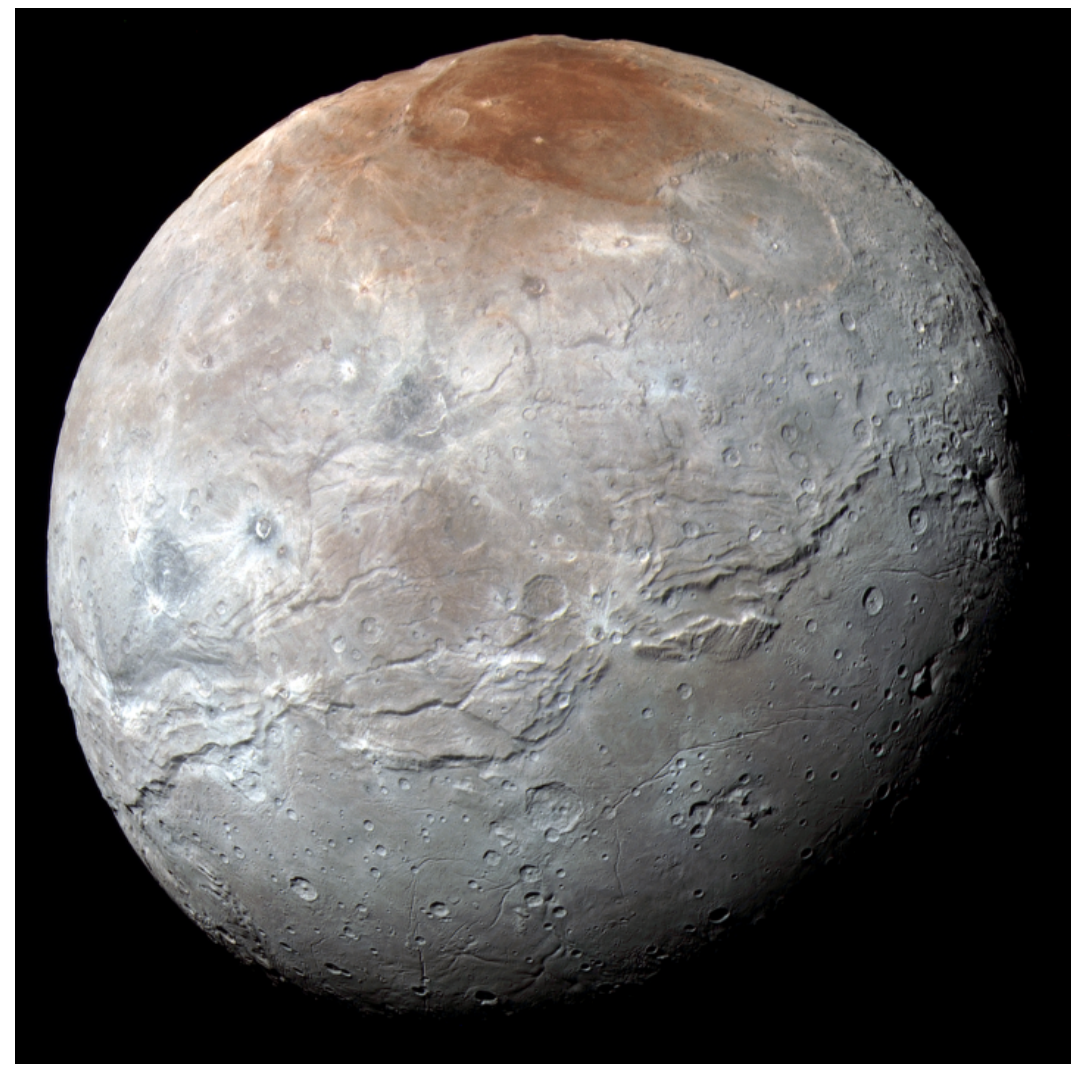

*Charon, featuring its red polar feature, ancient terrains, and massive tectonics. Credit: NASA / JHU-APL / SwRI / Alex H. Parker*

A striking feature of Charon is the clear presence of $NH_3$ (Dalle Ore *et al.*, 2018) inferred from a weak 2.2 μm band (which could be more effectively mapped at longer wavelengths than *New Horizons*' instrumentation was designed to reach). Exposed $NH_3$ is readily destroyed by radiolysis, so its abundance in certain crater ejecta could indicate recently exposed interior material (Grundy *et al.*, 2016). There could be a radiolytic cycle involving a more stable ammoniated molecule, or it could also be diffusing out from Charon's interior, all of which would be important for a future mission to investigate. Additionally, multiple regions of patterned ground in Vulcan Planitia—small pits at the ~100s m scale—hint at volatile escape (Robbins *et al*., 2019), offering clues to the chemical evolution of Charon's interior.

Furthermore, an ancient global expansion is also implied by the large polygonal blocks separated by deep graben in Oz Terra (Beyer *et al.*, 2017, 2018; Schenk *et al.*, 2018). These graben appear throughout much of the encounter hemisphere by >20 km of vertical relief and several multi-kilometer-deep canyons. Why the tectonic and cryovolcanic response differs in Oz Terra and Vulcan Planitia is unclear, and whether or not such expressions are also present elsewhere on the 60% of Charon not seen at high resolution by *New Horizons* compel a revisit.

## 2.4 Small Satellites and Satellite System Origin

Pluto's four tiny outer satellites were all discovered from Hubble Space Telescope observations in support of the *New Horizons* mission (Weaver *et al.*, 2016; Showalter *et al.*, 2011, 2012). They present a striking contrast to the giant moon, Charon, each being $10^6$–$10^8$ times less massive than Pluto's binary companion. The satellite system's coplanar, circular orbits indicate that it most likely originated from the Charon-forming giant impact (Stern *et al.*, 2006). Detailed numerical modeling of this process (Ward & Canup 2006; Kenyon & Bromley 2014; Walsh & Levison 2015) has been unsuccessful at reproducing the key orbital characteristics of the system, specifically that the moonlets orbit close to, but not directly in, the N:1 mean-motion resonances with Charon. The moons are also likely influenced by exotic three-body resonances (Showalter & Hamilton, 2015), which greatly complicate the orbital and perhaps also their spin/obliquity dynamics. The Pluto system thus presents key standing challenges—as well as opportunities—to understanding giant impact satellite formation and evolutionary dynamics.

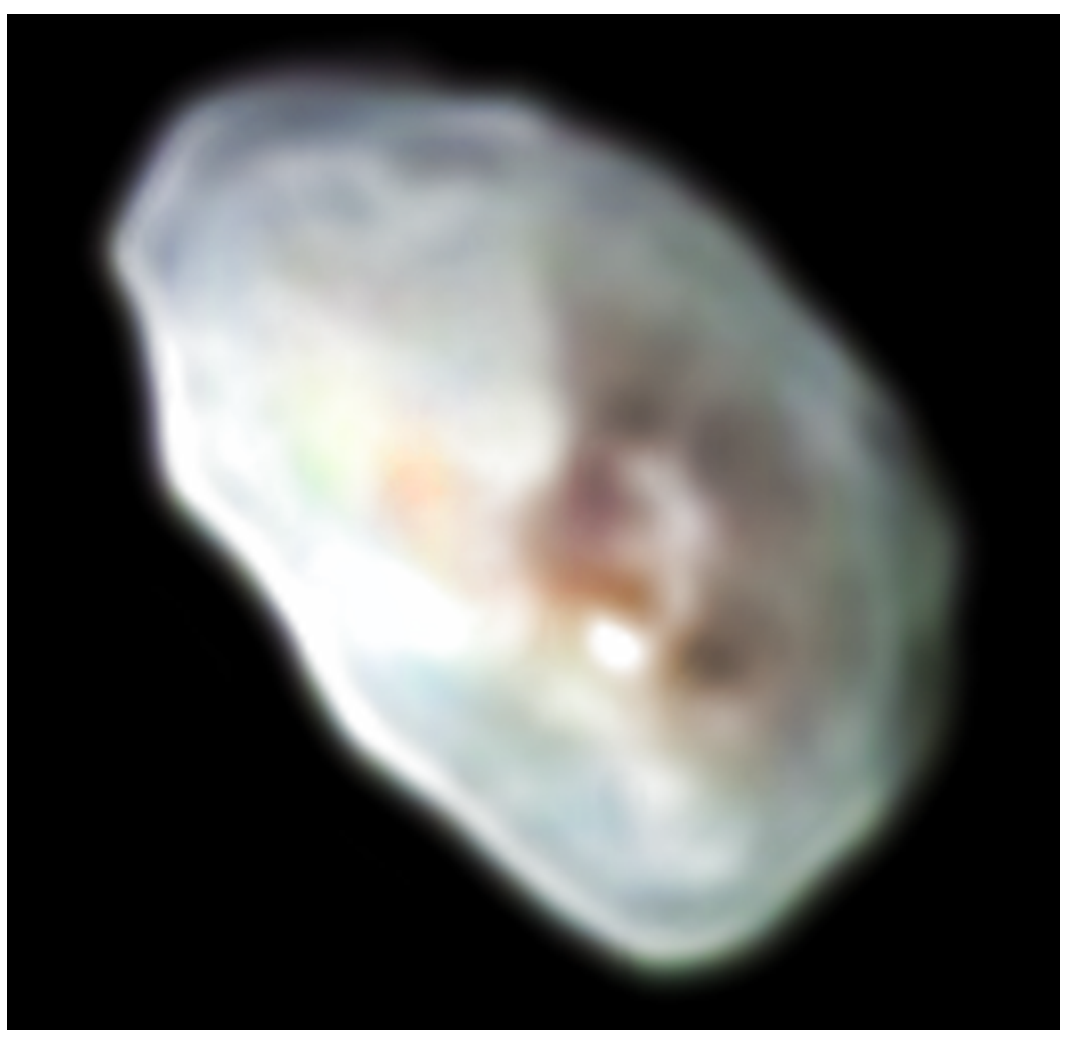

*Best Nix image, in stretched color. Credit: NASA / JHU-APL / SwRI*

The discovery of the uniformly high albedo, water-ice surfaces of the four small satellites, compared with similarly sized KBOs and even Charon, was surprising, as was the diversity of their bulk densities. Understanding the origin of these attributes and exploring the detailed geologies of these bodies is something *New Horizons* could not accomplish but which begs for future exploration. Comparative studies of these bodies to the cold classical KBO flyby of Arrokoth is also of extreme interest, but impossible without a revisit to study these small satellites at comparable resolutions to the *New Horizons* flyby of Arrokoth.

To better understand the small satellites and their origins, the next Pluto mission should include multiple close flybys of all four satellites to make detailed geological and compositional maps, direct measurements of their densities, optical studies of their regolith microphysical properties, and thermal mapping of their surfaces.

## 2.5 Pluto's Value to Further Understand the Kuiper Belt and the Kuiper Belt's Many Dwarf Planets

The exploration of the Pluto system by *New Horizons* was initially predicated on being the first reconnaissance of bodies in the Kuiper Belt, with Pluto being the largest, longest and best known, and most studied world therein (Belton, National Research Council 2003). The dynamical structure of the Kuiper Belt is the primary evidence for and the greatest test of hypotheses for an early dynamical instability/rearrangement of the Solar System (e.g., Levison *et al.*, 2008; Nesvorný *et al.*, 2016; and many others). All of these models share the inference that Pluto, lodged in a 3:2 mean-motion resonance with Neptune, formed in an ancestral planetesimal disk closer to the Sun, likely between 20 and 30 A.U. (Malhotra, 1993).

The Pluto system thus reflects the physical and chemical properties of this key region of the original solar nebula. The Pluto system is made even more valuable by the wide range of phenomena that it offers to teach us relevant to other Kuiper Belt dwarf planets. These include: satellite system formation, binary planet formation, and a wide range of Kuiper Belt planet surface, interior, and atmospheric processes.

## 3. The Logical Next Step in Pluto Exploration: Recommendations

Despite its size compared with Earth or Mars, Pluto is a world of extraordinary diversity, complexity and ongoing activity. Keys to its novel and active geology and geophysics, which operates in the absence of tidal heating, seem to include the prominent role of volatile ices, strong atmosphere-surface coupling, widespread tectonism, cryovolcanism, and a possible liquid interior ocean. At the frigid conditions of its surface, endogenic warmth from the decay of radioactive elements in its interior, as well as sunlight, are sufficient to mobilize ices such as $N_2$ and $CH_4$ in both vapor and solid form, and possibly as liquids in Pluto's past. Pluto could help answer how such novel and active processes work on bodies whose surfaces are dominated by volatile ices.

Pluto thus serves as the archetype for other dwarf planets of the Kuiper Belt, and it is the only Kuiper Belt world that is known well enough to justify a second-generation mission such as an orbiter or lander. Because none of the vexing problems opened by the *New Horizons* datasets are likely to be resolved from Earth or Earth orbit in the foreseeable decades, the case for returning to the Pluto system is strong.

As we have described here, the numerous, compelling, open scientific issues surrounding Pluto itself, the Pluto system in general, and the relationship of the Pluto system to the Kuiper Belt strongly motivate calls for follow on Pluto system exploration. Among the various options for that exploration are a second flyby, an orbiter, or a lander. Table 1 below compares these three options.

**Evaluating Table 1 (next page), we conclude that an orbiter is the best next step, and we support the findings that are discussed in the Persephone mission concept paper by Howett *et al*. (white paper submitted to this decadal survey).**

## 4. Equity, Diversity, and Inclusion

Any return to the Pluto system will take at least a decade and will encompass broad sections of planetary science, geoscience, astronomy, technology, engineering, and beyond. Studies of scientific teams have repeatedly demonstrated the importance of an integrated approach, where team members with diverse expertise develop synergies between their specialties and resources that result in an end product that adds up to more than the sum of its parts (Balakrishan *et al.*, 2011). Sociological studies have demonstrated that groups that foster strong connections across sub-units are more innovative (Burt, 2004; Powell & Koput, 1996; de Vaan *et al.*, 2015) with higher impact outcomes that endure (Curral *et al.*, 2001; de Vaan *et al.*, 2015).

Additionally, it is critical that the planetary science community fosters an interdisciplinary, diverse, equitable, inclusive, and accessible environment. We strongly encourage the Decadal Survey to consider the state of the profession and the issues of equity, diversity, inclusion, and accessibility—not as separable issues, but as critical steps on the pathway to understanding Pluto and the entire Solar System. Background information on the current lack of diversity in our community and specific, actionable, and practical recommendations can be found in several white papers that are planned to be submitted to this Decadal Survey (Rivera-Valentín *et al.*, 2020; Rathbun *et al.*, 2020; Strauss *et al.*, 2020; Milazzo *et al.*, 2020).

**Table 1: Possible Follow On Mission Architectures: Key Comparative Attributes**

| | **Flyby** | **Orbiter** | **Lander** |
|---|---|---|---|
| **Pros** | Lowest cost (New Frontiers Class) | Intermediate cost (Flagship Class) | Enables detailed high resolution surface process studies |
| | High TRL | High TRL | Enables seismic/heat flow studies and in-situ lower atmospheric and surface studies |
| | Shortest flight | Allows detailed full-system exploration | Possibly able to respond to new discoveries |
| | Map some unseen terrains on all bodies in the system, carry new instruments, look for temporal changes since *New Horizons* | Map all unseen terrains on all bodies in the system, carry new instruments, study temporal changes on many timescales, make in-situ upper atmospheric studies | Carry new instruments, study temporal changes |
| | Could also explore KBOs | Potentially explore KBOs beyond Pluto by leaving orbit | |
| | | Able to respond to new discoveries | |
| **Cons** | No extensive time variability studies | More expensive than a second flyby mission | Highest cost (Flagship Class) |
| | Not useful for many needed investigations (*e.g.*, altimetry, gravity) | Significant time requirements to reach the system with a low enough capture velocity, and the power and related age issues that result | Cannot go on to explore elsewhere in the Kuiper Belt |
| | | | Immature TRL |
| | | | Risky given current surface knowledge |
| | | | No landing site survey precursor |
| | | | Limited global or satellite studies |